# From SLA to Vendor-Neutral Metrics: An Intelligent Knowledge-Based Approach for Multi-Cloud SLA-based Broker

Víctor Rampérez, Javier Soriano, David Lizcano, Shadi Aljawarneh, Juan A. Lara .

*Abstract*—Cloud computing has been consolidated as a support for the vast majority of current and emerging technologies. However, there are some barriers that prevent the exploitation of the full potential of this technology. First, the major cloud providers currently put the onus of implementing the mechanisms that ensure compliance with the desired service levels on cloud consumers. However, consumers do not have the required expertise. Since each cloud provider exports a different set of low-level metrics, the strategies defined to ensure compliance with the established service-level agreement (SLA) are bound to a particular cloud provider. This fosters provider lock-in and prevents consumers from benefiting from the advantages of multi-cloud environments.

This paper presents a solution to the problem of automatically translating SLAs into objectives expressed as metrics that can be measured across multiple cloud providers. Firstly, we propose an intelligent knowledge-based system capable of automatically translating high-level SLAs defined by cloud consumers into a set of conditions expressed as vendor-neutral metrics, providing feedback to cloud consumers (intelligent tutoring system). Secondly, we present the set of vendor-neutral metrics and explain how they can be measured for the different cloud providers. Finally, we report a validation based on two use cases (IaaS and PaaS) in a multi-cloud environment formed by leading cloud providers. This evaluation has demonstrated that, thanks to the complementarity of the two solutions, cloud consumers can automatically and transparently exploit the multi-cloud in many application domains, as endorsed by the cloud experts consulted in the course of this research.

## I. INTRODUCTION

WE have seen how in just a few years society has transformed and evolved towards an increasingly digitalized world, where all aspects of its daily life depend on technology and more specifically on Internet services [1]. For all these reasons, it is not surprising that computer resources are now an essential utility in modern societies on a par with electricity, gas or water. As a result, cloud computing emerged as a service that provides computing resources as a service, that is, on-demand computing resources that users acquire on a pay-as-you-go basis.

Although several definitions of cloud computing have been proposed in the literature published by both academia and industry, most authors agree with the common elements present in the definition provided by the U.S. NIST (National Institute of Standards and Technology) [2]: *Cloud computing is a model for enabling convenient, on-demand network access to a shared pool of configurable computing resources (e.g. networks, servers, storage, applications, and services) that can be rapidly provisioned and released with minimal management effort or service provider interaction.*

Another important aspect of cloud computing is the service model. Cloud solutions can be divided into three categories according to their service model: (i) *Software as a Service* or *SaaS* in which users do not have control over the cloud infrastructure and therefore release their applications on a hosting environment (e.g. Gmail[1] or Google Docs[2]), (ii) *Platform as a Service* or *PaaS*, in which a platform supporting the full software lifecycle is provided for cloud consumers to develop cloud services and applications (e.g. AWS Elastic Beanstalk[3]) and (iii) *Infrastructure as a Service* or *IaaS*, where cloud consumers directly use IT infrastructures such as processing, storage or networks (e.g. Amazon EC2[4]).

In cloud computing, the service-level agreements (SLAs) are the contract between the cloud consumer and the cloud provider. They serve as the foundation for the expected level of service. As a result, SLAs are composed of objectives or service-level objectives (SLOs) expressed by means of high-level user guarantee parameters, such as throughput, cost or latency. SLOs are extremely important for meeting the expectations of all participants in order to assure business continuity, customer satisfaction and trust. They are, therefore, a critical long-term success factor. Nevertheless, there is a gap between SLO parameters, which are high-level metrics, and the metrics offered by the main cloud providers, which are usually low-level metrics, because these two metric levels are hard to understand and map. Currently, it is cloud consumers who are responsible for ensuring SLO achievement. This they do by adopting ad-hoc criteria, which raises serious issues [3]–[7]. In order to avoid serious service level deviations, we argue that the translation of SLAs should be automated, because cloud consumers neither know nor care about operation at such a low resource level.

V. Rampérez and J. Soriano are with Universidad Politécnica de Madrid (UPM), 28660 Boadilla del Monte, Madrid, Spain (e-mail: v.ramperez@upm.es, jsoriano@fi.upm.es).

D. Lizcano and Juan A. Lara are with Madrid Open University (UDIMA), 28400 Collado Villalba, Madrid, Spain (e-mail: david.lizcano@udima.es, juanalfonso.lara@udima.es).

S. Aljawarneh is with Jordan University of Science and Technology, JUST, 22110, Irbid, Jordan (e-mail: saaljawarneh@just.edu.jo)

[1] https://www.gmail.com
[2] https://www.google.es/intl/en/docs/about/
[3] https://aws.amazon.com/elasticbeanstalk/?nc1=h_ls
[4] https://aws.amazon.com/ec2/

In addition, many of the most important service levels demanded by cloud consumers in their SLAs (e.g. cost optimization, legal restrictions, high availability or vendor lock-in avoidance) cannot be adequately satisfied through the use of a single cloud provider. Therefore, it is necessary to use resources and services from multiple clouds. However, this poses an additional problem when mapping high-level metrics to low-level metrics, since there is no vendor-neutral set of metrics that can be monitored across multiple cloud providers [5], [7], [8].

The present work complements and extends our previous work [9], in which we presented FLAS (Forecasted Load Auto-Scaling), an auto-scaler for distributed services that combines the advantages of proactive and reactive approaches according to the situation to decide the optimal scaling actions in every moment. While the previous work addressed the problem of deciding the specific values of the dimensions of a scaling action (when to scale, how to scale and which resource to scale) by combining intelligent auto-scaling and reactive scaling models, this paper extends the previous work by allowing to exploit multi-cloud environments in a transparent way to cloud consumers through the use of intelligent computation.

This paper addresses and presents a solution to the problem of automatically translating SLAs into objectives expressed as metrics that can be measured across multiple cloud providers, both at the PaaS and IaaS level. To do this, we addressed two research questions tackling the two main problems mentioned above. This paper is structured around the statement of these problems as research questions, and their solution and validation. First, we address the problem of the automatic translation of SLAs into objectives described as metrics that can be measured in cloud environments. We argue that this translation is intrinsically dynamic due to the variability of the domain and application context. This rules out a static translation like a mapping. We therefore propose an intelligent knowledge-based system capable of performing this translation taking into account the context and application domain. Second, we address the problem of how to offer a set of metrics that can be measured across multiple cloud providers, abstracting the specificities of each cloud provider. To this end, we propose a harmonized and rather generic set of vendor-neutral metrics that can be measured in multiple clouds. Each of these cross-provider metrics would then have to be mapped to the corresponding metric in each cloud provider by means of mapping functions. Finally, we report a validation of the proposed solution through two use cases (IaaS and PaaS) in a multi-cloud environment formed by leading cloud providers.

The remainder of the paper is structured as follows. Section II sets out two research questions to further refine the problem that actually motivated this paper. In Section III, we give a brief review of the related work. Section IV provides a global overview of the proposed solution, which is developed in detail through the study and solution of RQ1 and RQ2 in Sections V and VI, respectively. Section VII shows the validation of the proposed solution through two real use cases with the main cloud providers. Finally, we highlight the findings of this research and state our future goals in Sections VIII and IX, respectively.

## II. PROBLEM STATEMENT

In recent years there has been an explosion of different cloud technologies, solutions and/or services that has led to different potential angles from which to address cloud computing concepts. This has blurred the boundaries between the different concepts. For the sake of clarity, we will use the following terminology from now on:

**Cloud consumer**: a person or organization that maintains a business relationship with, and uses the service from, a cloud provider [2]. It is the user of the services offered by the cloud provider (i.e. IaaS, PaaS or SaaS) [10], [11].

**Cloud provider**: a person, an organization, or an entity responsible for making a service available to cloud consumers [2]. It refers mainly to the IaaS, PaaS and/or SaaS provider [10], [11].

**Multi-cloud**: the usage of multiple and independent clouds by a client or a service. Multi-cloud and federated cloud are the two types of delivery models in multiple clouds. In the case of federated clouds, there is an agreement on resource sharing between the clouds involved, whereas no such agreement exists in the case of multi-cloud. In this paper, we adopted the multi-cloud model because it is hard to reach agreements between large cloud providers, especially since it is necessary to control the resources and services that an external provider expects to use for its benefit [8]. Additionally, multi-cloud differs from hybrid cloud in that it refers to multiple cloud services rather than multiple deployment modes (public, private or legacy).

**Broker**: a cloud broker is, according to the NIST Cloud Computing Reference Architecture [2], an entity that manages the use, performance, and delivery of cloud services, and negotiates relationships between cloud providers and cloud consumers. Cloud brokers provide a single point of entry for managing multiple cloud services and a single consistent interface to multiple differing providers.

As mentioned above, cloud computing SLAs are the contract between the cloud consumer and the cloud provider serving as the foundation for the expected level of service. The core of this contract is the specification of service level objectives (SLOs). SLOs are expressed by means of high-level user guarantee parameters such as throughput, availability, cost or latency. SLA compliance is extremely important to fulfill the expectations of all participants in order to assure business continuity, customer satisfaction and trust. It is, therefore, a critical long-term success factor [4]. However, the main cloud providers most widely used these days do not offer cloud consumers the possibility of defining their SLAs or agreeing on the service level that they expect from the cloud provider. These cloud providers merely offer a set of, usually very low-level, monitoring metrics for the services that they provide together with resource provisioning and management mechanisms. In fact, they put the onus of implementing the mechanisms that ensure compliance with the desired service levels on cloud consumers [5]. We argue that this is a misguided and error-prone approach because:

*Problem 1:* The metrics monitored and offered by cloud providers are too close to machine level and do not have the primitives for defining service-relevant metrics [12]. According to [3], there is a gap between monitored metrics, which are usually low-level metrics, and SLA agreements, which are high-level. In addition, mapping SLA parameters to low-level metrics exported by cloud providers is a tricky problem, which requires the knowledge of experts who understand the relationship between these two metric levels [6]. Therefore, by assigning this responsibility to cloud consumers that do not have the required expertise, and neither know nor care about how their applications behave at such a low level (since they are interested in high-level SLA metrics, like cost, latency, throughput or availability), causes serious problems for many organizations when they try to formulate their SLA strategies and adjust the metrics to support such strategies [4], [13].

In order to ensure compliance with some of the most demanded and important SLA service levels (e.g. cost optimization, legal restrictions or high availability), as they very often constitute the main reason for opting for cloud environments, it is necessary to use resources and services from multiple cloud providers, as a single cloud provider cannot offer these service levels. There are several reasons for opting for multi-cloud environments [8]: using external resources to deal with peaks in service or resource requests, optimize costs, adhere to constraints (e.g. new locations, laws or data sovereignty), achieve high availability by replicating the application or ensure back-ups to deal with scheduled inactivity or disasters. Therefore, it is not surprising that there is an increasing tendency[5] in recent years to use multiple clouds. However, the use of multiple clouds introduces several challenges due to the number of different cloud providers and their intrinsic variability. This translates into numerous technological and administrative barriers that are an obstacle to the use of multiple clouds, favoring vendor lock-in. This is a consequence of the lack of widely adopted standards, motivated by the reticence of the main cloud providers who regard such standards as posing a threat to innovation and market advantage rather than as a business need or priority [7], [8]. All of this introduces an additional problem related to Problem 1 mentioned above:

*Problem 2:* There is no vendor-neutral set of low-level metrics that can be monitored across cloud providers [5]. Since each cloud provider offers a different set of these low-level metrics, the strategies defined by cloud consumers to ensure SLA compliance are bound to a particular cloud provider, fostering provider lock-in and making it impossible to benefit from the service levels inherent to multi-cloud environments.

The novelty of our proposal lies in how we address and validate the following hypothesis. For this purpose, we state two more specific research questions, whose answers represent the solution to the abovementioned problems and constitute the backbone around which the remainder of the paper is structured:

*Hypothesis 1:* It is possible to automatically provide a cloud consumer-transparent translation of SLOs to conditions expressed as metrics that can be measured on multiple cloud providers, compliance with which also ensures adherence to the SLA.

- *RQ 1:* Is it possible to automatically establish the relationship between high-level objectives (or SLOs) and conditions expressed as low-level metrics for a given cloud application? (Problem 1)
- *RQ 2:* Is it possible to define a generic enough and extensible set of vendor-neutral and cross-provider measurable metrics? (Problem 2 )

## III. RELATED WORK

This section reviews the main contributions of both academic studies and major commercial solutions. More specifically, we first discuss studies that identify and highlight the importance of addressing the research questions motivating this research. We then review and analyze the different solutions existing in both industry and academia.

Many studies have identified the need to establish some kind of relationship or mapping between the high-level metrics in which cloud consumers are interested and the low-level metrics offered by cloud providers in order to establish mechanisms to ensure that the service levels demanded by cloud consumers are met (RQ1). According to [3], there is a gap between monitored metrics (low-level entities) and SLAs (high-level user guarantee parameters) and none of the approaches discussed in their work deal with the mappings of low-level monitored metrics to high-level SLA guarantees necessary in cloud-like environments. In the same vein, Paschke et al. [4] highlight the problem of the poor translation of SLAs into low-level metrics, claiming that the metrics used to measure and manage performance compliance to SLA commitments are the heart of a successful agreement and that inexperience in the use and automation of performance metrics causes problems for many organizations as they attempt to formulate their SLA strategies and set the metrics needed to support those strategies. Springs et al. [6] again also clearly identify the need to address this problem, stating that a key prerequisite for meeting these goals is to understand the relationship between high-level SLA parameters (e.g., availability, throughput, response time) and low-level resource metrics, such as counters and gauges. However, it is not easy to map SLA parameters to metrics that are retrieved from managed resources. Specifically domain experts are usually involved in translating these SLOs into lower-level policies that can then be used for design and monitoring purposes, as this often necessitates the application of domain knowledge to this problem. [14]

On the other hand, many authors have identified the need to establish a set of vendor-neutral metrics (i.e. they can be measured in multiple cloud providers) in order to offer a unified platform-independent set of metrics to monitor the cloud services, reducing the semantic interoperability between heterogeneous cloud providers (RQ2). For example, Patel et al. [5] argue that there is no universal set of metrics that can be monitored across cloud providers. They recognize that there are attempts to standardize the clouds and underscore the importance of such efforts in the light of monitoring capabilities. They also suggest a set of basic metrics and best practices for measurements be established. Quinton et al. [7] recommend using several clouds to deploy a multi-cloud configuration to deal with this problem, although they introduce several challenges due to the number of providers and their intrinsic variability. In the same vein, Ferrel et al. present a survey of cloud solutions and address the problem of heterogeneous clouds, claiming that the ability to run and manage multi-cloud systems can exploit the peculiarities of each cloud solution [15]. However, these cloud solutions are typically heterogeneous and the features that they provide are often incompatible. This diversity is an obstacle for the proper exploitation of the full potential of cloud computing.

At the time of writing, there were many commercial multi-cloud solutions available, including RightScale[6], Scalr[7] and Morpheus[8]. They all make it possible to deploy and manage applications in multiple clouds, acting as multi-cloud brokers. Cloud consumers can choose between a catalog of different types of instances (i.e. VM) defined in templates with different computing, RAM or storage capacities. Although cloud consumers can benefit from some of the features of a multi-cloud solution (high availability, cost optimization, geo-localization awareness, etc.), none of these commercial solutions address the issues that we raise in this paper. First of all, these solutions do not require a vendor-agnostic set of metrics, since they make use of solutions such as collectd[9] or Nagios[10], which offer a harmonized set of metrics. However, this approach limits the use of multi-cloud-to-IaaS services, since the installation of these monitoring frameworks requires access to the management of the instance or VM. Therefore, they are not able to exploit other high-value (e.g. PaaS-level) services from the main cloud providers. In addition, all of these commercial solutions follow, according to the classification by [16], a Trigger-Action approach, in which the cloud consumer specifies a set of triggers and their associated action(s) in order to ensure compliance with their SLAs. This approach of putting the onus of translating the SLAs into low-level conditions on cloud consumers has, as already mentioned, many drawbacks, as it requires expert knowledge about the application domain that they do not normally possess. To the best of our knowledge, there is currently no commercial solution offering a SLA-based multi-cloud broker service.

On the other hand, there are many academic projects and papers dealing with several of the above problems. We applied the systematic review approach proposed in [16], using the group taxonomy that it specifies. In this manner, we extended this broad survey, as shown in Table I. SALOON is a solution based on the combination of ontologies with feature models (FMs) to identify multi-cloud configurations whose functionalities and resources match the technical requirements of cloud applications [17]. It highlights the use of ontologies to formally capture and model the knowledge of cloud platforms and deal with the wide variety of terms used by different cloud providers to refer to the same concept, as a first step to reduce the semantic interoperability of multi-cloud environments. LoM2HiS is a novel framework for mapping low-level resource metrics to high-level SLA parameters [3]. This framework is embedded in the FoSII (Foundations of Self-Governing ICT Infrastructures) infrastructure for developing infrastructures for autonomic SLA management and enforcement [18]. To be precise, it detects SLA violation threats based on predefined threat thresholds and notifies the enactor component to avert the threats. Although it addresses the issue of mapping high-level metrics into low-level metrics, this solution is rather limited, as it is designed for only IaaS-level cloud services and not suitable for a multi-cloud environment, because monitoring requires the instrumentalization of instances. Similarly, the mOSAIC project [19] offers an open source API and platform for the development and deployment of applications in multi-cloud environments. For this purpose, it is possible to define the service requirements of an application, at both the IaaS and PaaS levels, in terms of a cloud ontology, with the platform being responsible for finding the services that best match the requirements of the application. Nonetheless, this solution does not offer a multi-cloud service at the PaaS level. Instead, it requires the customization of IaaS resources to meet PaaS requirements and does not address the problem of the gap between high- and low-level metrics. Cloud4SOA is a solution for multi-cloud application management, monitoring and migration at the PaaS level [20]. More specifically, it offers a multi-cloud broker that addresses the challenges of semantic interoperability at the PaaS level in public, private and hybrid deployment models. In order to monitor the heterogeneous PaaS resources offered by a multi-cloud environment and avoid vendor lock-in, Zeginis et al. propose a set of harmonized metrics and adapters that bridge the business logic between the harmonized API and the PaaS offerings. However, the set of harmonized vendor-neutral metrics that are proposed are very limited. Therefore, it is not possible to define cloud consumer SLAs in terms of these metrics.

We agree that a solution that addresses the above problems and solves all the research questions raised would be a more valuable real contribution than the commercial and academic partial solutions presented above. To the best of our knowledge, there is no such solution. This is the main contribution of this research.

---

[6] https://www.rightscale.com/
[7] https://www.scalr.com
[8] https://www.morpheusdata.com
[9] https://collectd.org/
[10] https://www.nagios.org/

TABLE I: Summary of academic cloud brokering solutions

| | Organization Type | Architecture | Brokering Type | Application Type | Awareness | Cloud Services |
|---|---|---|---|---|---|---|
| **SALOON** | Ministry of Higher Education and Research, Nord–Pas de Calais Regional Council, the FEDER through the Contrat de Plan Etat Region Campus Intelligence Ambianteand the EU FP7 PaaSage project. | Independent Service | SLA based | Singular jobs, periodical jobs, compute-intensive and data-intensive interactive application | Pricing, location | PaaS IaaS |
| **LoM2HiS** | Vienna Science and Technology Fund under grant agreement Foundations of Self-governing ICT Infrastructures (FoSII) | Independent Service | SLA based | Singular jobs, periodical jobs, compute-intensive and data-intensive interactive application | - | IaaS |
| **mOSAIC** | Private and Public European research organizations funded by EU | Independent Service | SLA based | Singular jobs, periodical jobs, compute-intensive and data-intensive interactive application | Pricing | PaaS IaaS |
| **Cloud4SOA** | Partially funded by the European Commission in the context of the Cloud4SOA project | Independent Service | SLA based | Singular jobs, periodical jobs, compute-intensive and data-intensive interactive application | Pricing | PaaS |

## IV. PROPOSED SOLUTION OVERVIEW

The aim of this section is to provide a brief overview of the solution proposed here, focusing on the stated research questions and the synergies arising out of their joint response and also discussing issues that are out of scope.

As shown in Figure 1, the proposed solution focuses on the cloud consumer viewpoint. As already mentioned, cloud consumers use cloud environments (and especially multi-cloud) in order to satisfy the requirements and service levels of their applications. For this reason, it is considered that the solution should be designed based on the SLA. The SLA is where cloud consumers state these needs, which are a key long-term success factor. SLAs are composed of a set of service-level objectives or SLOs, which are SLA parameter (slap) or high-level metric conditions. These conditions are very wide ranging, including, for example, throughput greater than 30 k notifications/s, all resources deployed in USA regions (due to restrictions or data sovereignty laws), monthly cost never greater than $100, etc. This work does not deal with the formalization of SLAs and SLOs, since there are many works offering frameworks and standards valid for this purpose [21].

These high-level metric conditions (SLOs) are then translated into lower-level metric conditions (RQ1). This translation cannot be limited to static mapping, as the relationship between SLA parameter conditions and measurable metrics has to take into account numerous contextual factors, such as the application domain. The following section focuses on describing the proposed solution to this problem. In addition, we propose a repository of metrics that can be measured across multiple cloud providers, which does not depend on any specific cloud provider (RQ2). This vendor-neutral cross-provider measurable metrics repository is divided into IaaS, PaaS and Context metrics (e.g. geo-location, cost, or energy consumption metrics). In this case, the relationship between this vendor-neutral metric repository and the metrics exported by each cloud provider has been established through static mapping functions, as explained in more detail later. These mapping functions, which bridge the gap between the harmonized standard repository of metrics and the metrics exported by each cloud provider, will be offered by a cloud provider-specific adapter implementing a common API exported by the vendor-neutral metrics repository. These adapters are also responsible for communicating with the corresponding cloud provider. To do this, they use one of the multiple inter-cloud libraries available for various programming languages (e.g. jclouds[11] for JAVA or Apache LibCloud for Python[12]), which have been widely tested through use in multiple multi-cloud projects [19]. These libraries abstract the heterogeneity of the different cloud providers' management APIs and allow resource provisioning.

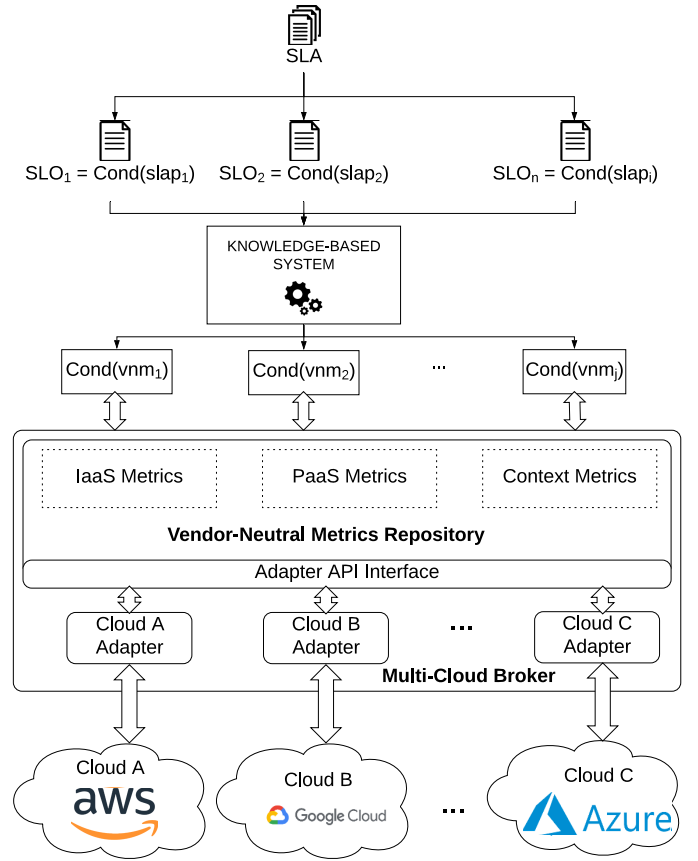

Fig. 1: Overview of the proposed solution focused on the vision of the cloud consumer. The top section shows how the SLA and its respective SLOs are converted into a set of low-level conditions, which are expressed as vendor-neutral metrics that can be monitored in multiple-cloud environments.

As a result, the proposed solution can ensure compliance with the service levels required by cloud consumers by abstracting the specificities between the services offered by multiple cloud providers. This provides cloud consumers with a high level of specification of their objectives, which can be evaluated in a standardized and cross-provider manner.

There are many other challenges that a multi-cloud broker must address (e.g. provisioning, allocation and portability of resources or conditions enforcement). For the sake of clarity, they have not been included in this paper, as the novelty of this research resides in the innovative solution to the two research questions. However, these aspects are orthogonal to this research, which could be rounded out by integrating some of the diverse solutions to each of these challenges proposed in the literature.

## V. FROM SLA TO LOW-LEVEL CONDITIONS (RQ1)

This section describes the solution that we propose to deal with RQ1 above and the empirical grounds for adopting this solution. As previously explained, the solution to RQ1 implies offering an intelligent knowledge-based system that is capable of automatically determining the low-level metric conditions (i.e. key performance indicators or KPI) necessary to achieve a high-level metric condition in order to ensure compliance with the high-level metric objective (SLO).

The relationship between high-level objectives or SLOs and low-level objectives raise the following needs, which reflect the requirements that a solution to RQ1 must satisfy. It must:

*N 1:* Capture and use expert knowledge. The translation of high-level objectives or SLOs into their corresponding low-level objectives is a difficult problem and requires the application of expert knowledge [14]. Therefore, the proposed system must be able to capture and use this expert knowledge to decompose SLOs into their respective low-level metric conditions or KPIs.

*N 2:* Be flexible and adaptive to changes. Because SLAs are defined at a very high level, there is a gap between these agreements and the underlying reality, which changes enormously and very rapidly. For example, a user wants the resources supporting the contracted cloud services to be in a particular region (e.g. Europe) in order to comply with the applicable data laws. However, the data center where these resources are hosted may, in the future, be in a country that no longer belongs to the region that complies with these regulations. For example, "The local laws that apply in the jurisdiction where the content is located are an important consideration for some customers. However, customers also need to consider whether laws in other jurisdictions may apply to them" [22]. Another example of these possible changes is the variation in the pricing policy of a given cloud provider, which may compromise the assurance of complying with a given cost-related SLO. Since these changes are unpredictable and inherent to our reality, the proposed system must be able to adapt to these changes, if not automatically, with the least possible change, favored by a design conceived to deal with this variability.

*N 3:* Offer dynamic mapping. The relationship between high-level and low-level conditions cannot be expressed by a simple static mapping, as this relationship is strongly influenced by the context and specific domain of each case. However, there are two key factors to be taken into account when establishing these relationships, which therefore must be considered by the proposed solution. The first factor is the application type, since the same SLO cannot generally be expressed by the same low-level metric conditions for different types of applications built for different purposes. For example, an application throughput SLO will be defined by different low-level metric conditions for a database-type application (whose limiting factor may be storage) than for publish/subscribe middleware (whose limiting factor may be processing capacity or bandwidth). The second factor is the application domain, which accounts for the fact that the same application can perform very differently at resource level depending on the application domain. This obviously has to be considered to ensure compliance with the SLOs defined by the cloud consumer. Figure 2 shows how four middleware applications, implementing the same publish/subscribe paradigm, have different resource-level performance depending on whether the model used to determine the matching between a subscription and a notification is based on a predefined topic (i.e. topic-based publish/subscribe or TBPS) or on the content of the message (i.e. content-based publish/subscribe CBPS). More specifically, two of the middlewares implemented the topic-based model (RabbitMQ [23] and ActiveMQ [24]) and the other two implemented the content-based model (E-SilboPS [13] and BE-Tree [25]). The experiment measures the throughput (notifications/s) of each of the middlewares with a publisher sending notifications at maximum rate to $n$ previously loaded subscribers in order not to influence the matching time. This study shows that, in the case of content-based brokers (E-SilboPS and BE-Tree), the throughput saturation point is due to CPU exhaustion (i.e. they are CPU-bound applications). On the contrary, the resource that limits the throughput in the case of topic-based brokers (RabbitMQ and ActiveMQ) is memory (i.e. they are I/O-bound applications). This goes to show that systems implementing the same paradigm can perform very differently at resource level depending on the purpose for which an application is to be used. As a result, the same high-level metrics (throughput) may require different low-level metric mapping functions depending on the task KPIs, since, as we have seen, it is the KPIs (CPU usage and memory usage for the CBPS and TBPS middlewares, respectively) that determine the saturation point of the high-level metric, and this could lead to non-compliance with the SLO. There are many similar examples to confirm that the application domain influences the low-level performance of the same application, such as the difference in the resource-level performance of the same relational database type application depending on whether it is used in an OLTP (on-line transaction processing) or OLAP (on-line analytical processing) context.

In order to meet all the above requirements, we propose an intelligent knowledge-based system as shown in Figure 4. Intelligent knowledge-based systems use knowledge and inference procedures to solve problems that are so complex as to require human experience [26]. Besides, problem complexity again calls for best instead of optimal matching in most cases [8]. The system knowledge base should represent the expert knowledge required to translate SLOs into specific objectives to be met by low-level metrics. To do this, the production rules that form the knowledge base can use all the information available to provide the system with the necessary expressiveness to represent the expert knowledge (N1). Figure 3 formally shows the structure of the production rules that populate the knowledge base. In addition, this production rule reflects part of the expert knowledge extracted from the evaluation results illustrated in Figure 2 for the case of CBPS middlewares.

Section VII shows the operation of this intelligent knowledge-based system in two real use cases that demonstrate how this solution satisfies all the abovementioned needs. However, the use of an intelligent knowledge-based system not only fulfills the abovementioned needs and is therefore a valid solution to the RQ1, but it also offers other advantages that add significant value. Intelligent knowledge-based systems can discover new knowledge. For instance, a correct classification and hierarchy of application types can lead to the discovery of new relationships between low-level and high-level metrics and between SLO and low-level metric objectives. In addition, this is highly flexible, as the inclusion of new rules can lead to a substantial system improvement. The source of this new knowledge acquisition is either: (1) a feedback loop that modifies or adds information to the knowledge base in the event of deviations in the SLOs or (2) experts who voluntarily expand or refine the knowledge base. Another advantage is that intelligent knowledge-based systems offer the opportunity to aggregate knowledge, allowing all cloud consumers to benefit from that expert knowledge. For example, many of the cloud applications follow a similar architecture (e.g. back-end and front-end with a database) benefiting from the production rules that have been found to perform such translations successfully. Moreover, when the application to be deployed is experimental as opposed to a solution widely used by the community (e.g. mogoDB[13]) whose low-level performance is quite well known, the use of an intelligent knowledge-based system can be very useful since intelligent knowledge-based systems are tolerant to and can handle uncertainty or missing data. Finally, this intelligent knowledge-based systems provide an explanation of the translations performed (intelligent tutoring system), which gives cloud consumers extra insight into the relationship between the performance of their systems and their objectives as expressed in the SLAs.

For the sake of clarity and simplicity, we use the most basic type of intelligent knowledge-based system to describe the proposed solution. However, other types of intelligent knowledge-based systems incorporating neural networks and fuzzy logic problem solving approaches [27] or different types of inference engines that are possibly much more suitable for specific contexts, could be used.

## VI. TOWARDS A SET OF VENDOR-NEUTRAL CROSS-PROVIDER MEASURABLE METRICS (RQ2)

This section presents and explains the solution proposed to solve RQ2, highlighting its advantages and its justification through some examples of its operation that are extended in the real use cases presented in the evaluation section. As mentioned above, the answer to RQ2 is to offer a harmonized set of monitoring metrics, which aims to abstract the specificities between multiple cloud provider offerings, making the monitoring and enforcing of SLOs in a multi-cloud environment transparent for cloud consumers. This set of vendor-neutral and cross-provider measurable metrics will be used by the solution to RQ1 explained above to express the low-level conditions that ensure the enforcement of SLOs (see Section V).

Many authors note that there is no set of metrics that can be monitored for all cloud providers [5], [7], [15]. To solve this and offer a solution to RQ2, we proposed a repository of cloud provider-agnostic metrics (vendor-neutral metrics or vnm) that can be monitored for the main cloud providers most commonly used today. This vendor-neutral metrics repository and the mapping functions that relate each of the metrics to those exported by each cloud provider provide a solution for abstracting the specificities of each cloud provider, offering a set of provider-agnostic and measurable metrics in multi-cloud environments. In addition, the set of metrics offered by this repository is very large and diverse, ranging from IaaS metrics, through PaaS metrics, to context metrics. This variety and number of metrics perfectly complements the above solution to RQ1, since the wider the variety of metrics, the more optimal the translations of SLOs to vendor-neutral metric objectives will be. In the following sections, we explain the proposed vendor-neutral metrics and how they can be built in order to be cross-provider measurable for IaaS, PaaS and context.

### A. IaaS metrics

The evident heterogeneity of the metrics exported by the main cloud providers has led the major multi-cloud services and brokers to consider alternative solutions to this problem. The instrumentalization of IaaS resources (i.e. VM or

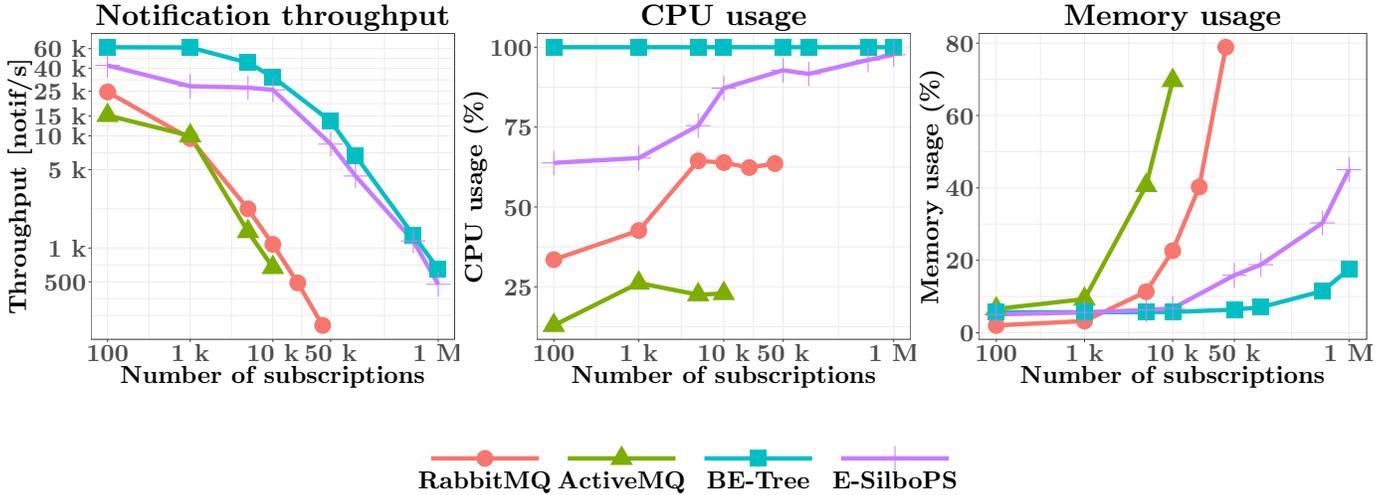

Fig. 2: Four publish/subscribe middleware applications widely used in cloud environments, showing that the throughput saturation point in content-based publish/subscribe middlewares or CBPS (i.e. E-SilboPS and BeTree) is due to CPU exhaustion, whereas it is due, in the case of topic-based publish/subscribe or TBPS (i.e. ActiveMQ and RabbitMQ), to lack of memory.

**IF** $AT = CBPS \land throughput \geq 10knotif/s$

**THEN** $CPU\_usage \leq 80$

Fig. 3: Example of a production rule reflecting the expert knowledge about the relationship between throughput (high-level metric) and its respective KPI, which is CPU usage (low-level metric) for the case of content-based publish/subscribe (CBPS) middlewares.

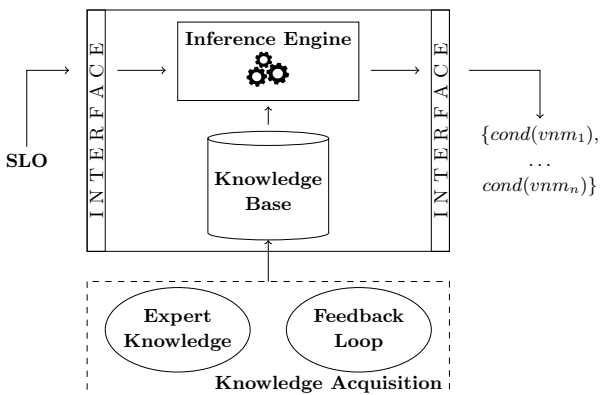

Fig. 4: The intelligent knowledge-based system converts SLOs into a set of low-level metric conditions, using an inference engine fed by a knowledge base populated with expert knowledge and feedback from previous translations.

instances), using tools like collectd[14] or Nagios[15], which export a harmonized set of metrics independent of the instance provider, has been the most adopted solution. Some cloud providers like AWS even require cloud consumers to instrumentalize their instances through the use of scripts for more detailed monitoring (since otherwise it is impossible to get metrics such as RAM usage[16]). For this reason, we argue that, in the case of IaaS, the use of tools such as Nagios or collectd is the best possible solution to RQ2, as these tools avoid the vendor lock-in caused by these proprietary instrumentalization scripts, offering a vendor-neutral set of low-level metrics. Collectd, in particular, has been used as a tool for instrumentalizing IaaS resources, where the metrics exported by collectd are the IaaS vendor-neutral metrics of the proposed repository, some of which are shown in Table II.

### B. PaaS metrics

The approach proposed for IaaS is not valid in the case of PaaS, since PaaS solutions do not usually offer access to the underlying infrastructure resources, such as VM or instances. On the other hand, some multi-cloud PaaS solutions do use this approach since the actually provide a false PaaS by merely using IaaS services on top of the software installation services necessary to offer PaaS solutions. Nevertheless, although this solution benefits from instrumentalization (not only at resource level as before, but also by exporting a vendor-neutral set of higher level metrics), it forgoes all the advantages of real PaaS (service optimization, lower cost, vendor lock-in avoidance and release from low-level resources management). These limitations are responsible for the small number of PaaS multi-cloud solutions and the fact that existing solutions are not real multi-cloud PaaS.

Firstly, we analyzed the services offered by the three main PaaS cloud providers (i.e. Amazon Web Services, Google

---

[14]https://collectd.org/
[15]https://www.nagios.org/

[16]https://docs.aws.amazon.com/AWSEC2/latest/UserGuide/mon-scripts.html

TABLE II: Some examples of the proposed IaaS vendor-neutral metrics

| Metric Type (plugin) | Metric |
|---|---|
| CPU | cpu-#.cpu-idle.value |
| | cpu-#.cpu-interrupt.value |
| | cpu-#.cpu-nice.value |
| | cpu-#.cpu-softirq.value |
| | cpu-#.cpu-steal.value |
| | cpu-#.cpu-system.value |
| | cpu-#.cpu-user.value |
| | cpu-#.cpu-wait.value |
| Disk | disk-<dn>.disk_merged.read |
| | disk-<dn>.disk_merged.write |
| | disk-<dn>.disk_octets.read |
| | disk-<dn>.disk_octets.write |
| | disk-<dn>.disk_ops.read |
| | disk-<dn>.disk_ops.write |
| | disk-<dn>.disk_time.read |
| | disk-<dn>.disk_time.write |
| Memory | memory.memory-buffered.value |
| | memory.memory-cached.value |
| | memory.memory-free.value |
| | memory.memory-used.value |
| Interface | interface-<int>.if_errors.rx |
| | interface-<int>.if_errors.tx |
| | interface-<int>.if_octets.rx |
| | interface-<int>.if_octets.tx |
| | interface-<int>.if_packets.rx |
| | interface-<int>.if_packets.tx |
| Other | uptime.uptime.value |

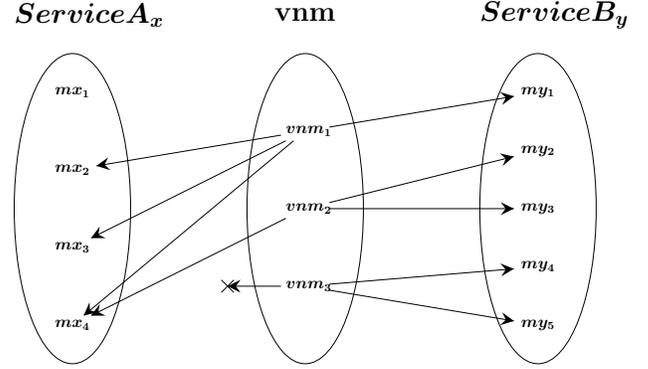

Fig. 5: In the figure, $ServiceA_x$ and $ServiceB_y$ are the implementation of an equivalent service in a cloud provider $x$ and $y$, and the metrics of these services $mx_i$ and $my_j$ respectively. The set of vendor-neutral metrics ($vnm$) is formed by all those metrics that can be measured in all the equivalent services of the different cloud providers, either directly ($vnm_1 = my_1$) or as a function of several metrics ($vnm_2 = f(my_2, my_3)$). If a $vnm$ can not be measured in one cloud provider, this metric has not been considered ($vnm_3$).

Cloud and Microsoft Azure). The conclusion of this analysis was that they mainly offer solutions for the same services. We derived the mapping between these equivalent solutions, shown in Table III, following the service mappings offered by the cloud providers themselves as documentation[17] for cloud consumers to migrate their services from other providers.

As shown in Figure 5, for each service, the metrics exported by the corresponding solution of each cloud provider were then analyzed. This analysis corroborated that all major cloud providers export similar monitoring metrics for each comparable service they offer, as they all measure the same magnitudes. There are slight differences between the metrics exported by each provider, the most common of which are differences in name, unit of measure (e.g. percentage or absolute value), sampled period, or composition of metrics (i.e. some cloud providers offer composite metrics that can be calculated as functions of simpler metrics exported by other cloud providers). On the one hand, we classified mappings where the relationships between the metrics exported by the services is established directly (metrics that measure the same magnitudes) or require negligible transformation (e.g. simple operations combining several directly exported metrics) as direct (mapping column of Table III). On the other hand, mappings that, due to the complexity of the service, require experts for their definition, which usually requires defining custom metrics based on very in-depth knowledge of the service, are classed as indirect. For the sake of clarity, this paper focuses on direct mappings, which are in actual fact the most demanded multi-cloud services. Nevertheless, we consider that the provision of these mappings by different service experts would be very valuable for the solution proposed in this paper. This could by done by engaging cloud providers in a similar manner as they are currently involved in providing migration guides between services from different providers. Finally, a small percentage of the solutions offered by cloud providers either do not provide a set of metrics for the respective service or implement the delivered metrics in a significantly different manner from other providers. In these cases, it is not possible to establish a relationship between the metrics of the different services, as indicated by the not possible value in the mapping column of Table III.

Some examples[18] of these PaaS vendor-neutral metrics and the related metrics exported by the equivalent solutions of the different cloud providers are shown in Tables IV, V, VI and VII for the block storage, object storage, PostgreSQL and MySQL database management system services, respectively. These tables show, for each service, the set of vendor-neutral metrics, their units of measurement and the metrics used for their calculation in each of the equivalent solutions of the different cloud providers. These relationships can be translated to the metrics exported by the cloud provider for a given service that are necessary to build the corresponding vendor-neutral metrics. More specifically, both metrics explicitly exported by the provider monitoring systems (e.g. Amazon Cloud Watch), and implicitly exported metrics, such as the geographical region where a service is deployed, the amount of RAM, the number of hard disks, etc., which are normally obtained from the service deployment configuration are used. For example, for the PaaS MySQL service offered by the three providers, Table VII defines a vendor-neutral metric called *memory_utilization*. This metric indicates the percentage of RAM memory being used. Both Google Cloud and Microsoft Azure export this metric directly. In the case of AWS, however, this metric can be calculated using the *FreeableMemory* (explicitly exported) and *TotalMemory* (service configuration-based) metrics.

TABLE III: Equivalence of main cloud provider services and their mapping relationship

| Service Category | Service | AWS | Google Cloud | Microsoft Azure | Mapping |
|---|---|---|---|---|---|
| Compute | IaaS | Elastic Compute Cloud | Compute Engine | Virtual Machines | Direct [*] |
| | PaaS Framework | Elastic Beanstalk | App Engine | App Service, Cloud Services | Indirect |
| | Containers | Elastic Container Service, Elastic Kubernetes Service | Container, Google Kubernetes Engine | Azure Container Service, Azure Service Fabric | Direct |
| | Serverless Functions | Lambda | Cloud Functions | Functions | Indirect |
| Network | Load Balancer Application | Elastic Load Balancer | Cloud Load Balancing | Application Gateway | Direct |
| | Load Balancer Networks | Elastic Load Balancer | Cloud Load Balancing | Azure Load Balancer | Direct |
| | Dedicated Interconnect | Direct Connect | Cloud Interconnect | ExpressRoute | Direct |
| | Domains and DNS | Route 53 | Google Domains, Cloud DNS | Azure DNS | Not possible |
| | CDN | CloudFront | Cloud Content Delivery Network | Azure CDN | Not possible |
| Storage | Object Storage | Simple Storage Service | Cloud Storage | Azure Blob Storage | Direct |
| | Block Storage | Elastic Block Store | Persistent Disk | Disk Storage | Direct |
| | Reduced-availability Storage | S3 Standard-Infrequent Access, S3 One Zone-Infrequent Access | Cloud Storage Nearline | Azure Cool Blob Storage | Direct |
| | File Storage | Elastic File System | Cloud Filestore (beta) | Azure File Storage | Indirect |
| Database | RDBMS | Relational Database Service | Cloud SQL | SQL Database | Direct |
| | NoSQL: Key-value | DynamoDB | Cloud Datastore, Cloud Bigtable | Table Storage | Indirect |
| | NoSQL: Indexed | SimpleDB | Cloud Datastore | Cosmos DB | Not possible |
| Big Data & Analytics | Batch Data Processing | Elastic MapReduce | Cloud Dataproc | Batch | Direct |
| | Stream Data Processing | Kinesis | Cloud Dataflow | Stream Analytics | Direct |
| | Stream Data Ingest | Kinesis | Cloud Pub/Sub | Event Hubs, Service Bus | Indirect |
| | Analytics | Redshift, Athena | BigQuery | Data Lake Analytics, Data Lake Store | Indirect |
| Application Services | Messaging | Simple Notification Service, Simple Queueing Service | Cloud Pub/Sub | Service Bus | Indirect |

[*] It is possible, but the preferred option is to use collectd, Nagios or similar.

This set of vendor-neutral metrics is only a partial solution to RQ2, since they should also be able to be monitored for all cloud providers. To this end, it is necessary to formally define how these provider-independent metrics can be calculated based on the metrics exported by each of the cloud providers. In other words, for every vendor-neutral metric defined in the repository, there is a mapping function that relates this metric to one or more metrics exported by each of the supported cloud providers. These functions establish a static mapping between the vendor-neutral set of metrics in the repository and the set of metrics exported by each cloud provider. In this case, however, this static mapping is a valid solution, as these relationships are not conditioned by highly variable external factors. This is ensured by the contract that represents the API used by each cloud provider to export its metrics and on which not only multi-cloud solutions (which could lead to a conflict of interest), but also a large number of monitoring solutions used by its own customers depend.

Continuing with the previous example, the vendor-neutral metric *memory_usage* of the MySQL service has a direct mapping function with the *database/memory/utilization* and *memory_percent* metrics of the equivalent Google Cloud and Microsoft Azure services, respectively, since they measure the

$$memory\_utilization =$$
$$= \left(1 - \left(\frac{FreeableMemory}{2^{30} \times TotalMemory}\right)\right) \times 100 \qquad (1)$$

Figure 6: Mapping function that relates the *memory_utilization* vendor-neutral metric to the equivalent AWS metrics.

same magnitude, with the same unit and the same sampling period. However, in the case of AWS, this vendor-neutral metric can be calculated by composing two metrics as indicated by the mapping function shown in Figure 6.

Another example that illustrates this type of mappings is the case of the vendor-neutral metric *io_operations* of the block storage service, which measures the number of input/output operations per second performed. As shown in Table VIII, this metric is calculated in the case of AWS by adding the *VolumeReadOps* and *VolumeWriteOps* metrics. As these metrics are collected every minute, however, it is necessary to divide by 60 to calculate the operations per second. In the case of Google Cloud, this is calculated by dividing the number of accumulated input/output operations by the accumulated input/output operations time. Finally, this metric is calculated in Microsoft Azure as the sum of all the input/output operation metrics for both data and OS, since the vendor-neutral metric unit measures them all.

However, these mappings should be thought of as resembling the process of an ETL (extract, transform and load) rather than merely in terms of mathematical functions. For example, if the metric sampling period is different for two cloud providers or if a metric is exported as an increment by one provider and as an accumulated value by another (Table IX), it is necessary to use complex event processing (CEP) solutions to establish a time-order window to calculate this type of values. As shown in Table IX, for example, the vendor-neutral metric *total_requests* represents the cumulative value of requests, which matches the metrics directly exported by the equivalent AWS and Microsoft Azure services. However, the Google Cloud solution exports this metric as the differential increment with respect to the previous value. Therefore, it is necessary to calculate the accumulated value of this metric.

TABLE IV: Block Storage Vendor-Neutral Metrics

| VNM | Unit | AWS Related metric(s) | GCP Related metric(s) | M. Azure Related metric(s) |
|---|---|---|---|---|
| disk_read | Bytes/s | VolumeReadBytes | read_bytes_count | Data Disk Read Bytes/Sec OS Disk Read Bytes/Sec |
| disk_write | Bytes/s | VolumeWriteBytes | write_bytes_count | Data Disk Write Bytes/Sec OS Disk Write Bytes/Sec |
| queue_length | count | VolumeQueueLength | pending_operations | Data Disk QD OS Disk QD |
| io_operations | operations/s | VolumeReadOpsVolumeWriteOps | operation_countoperation_time | Data Disk Read Operations/Sec Data Disk Write Operations/Sec OS Disk Read Operations/Sec OS Disk Write Operations/Sec |

TABLE V: Object Storage Vendor-Neutral Metrics

| VNM | Unit | AWS Related metric(s) | GCP Related metric(s) | M. Azure Related metric(s) |
|---|---|---|---|---|
| objects | count | NumberOfObjects | storage/object_count | BlobCount |
| total_requests | count | AllRequests | api/request_count | Transactions |
| data_downloaded | Bytes | BytesDownloaded | network/sent_bytes_count | Egress |
| data_uploaded | Bytes | BytesUploaded | network/received_bytes_count | Ingress |
| storage_used | Bytes | BucketSizeBytes | storage/total_bytes | BlobCapacity |
| get_requests | count | GetRequests | api/request_count | Transactions |
| put_requests | count | PutRequests | api/request_count | Transactions |
| delete_requests | count | DeleteRequests | api/request_count | Transactions |
| head_requests | count | HeadRequests | api/request_count | Transactions |
| post_requests | count | PostRequests | api/request_count | Transactions |
| 4xx_errors | count | 4xxErrors | api/request_count | Transactions |
| 5xx_errors | count | 5xxErrors | api/request_count | Transactions |

TABLE VI: PostgreSQL Vendor-Neutral Metrics

| VNM | Unit | AWS Related metric(s) | GCP Related metric(s) | M. Azure Related metric(s) |
|---|---|---|---|---|
| CPU_utilization | % | CPUUtilization | database/cpu/utilization | cpu_percent |
| memory_usage | Bytes | FreeableMemory (Bytes) TotalMemory (GiB) | database/memory/usage | memory_percentmemory_quota (GB) |
| memory_utilization | % | FreeableMemory (Bytes) TotalMemory (GiB) | database/memory/utilization | memory_percent |
| disk_usage | Bytes | FreeStorageSpace (Bytes) TotalDisk (GiB) | database/disk/bytes_used | storage_used |
| disk_utilization | % | FreeStorageSpace (Bytes) TotalDisk (GiB) | database/disk/utilization | storage_percent |
| disk_write_ops | Count/s | WriteIOPS | database/disk/write_ops_count | Data Disk Write Operations/Sec OS Disk Write Bytes/Sec |
| disk_read_ops | Count/s | ReadIOPS | database/disk/read_ops_count | Data Disk Read Operations/Sec OS Disk Read Bytes/Sec |
| network_received | Bytes/s | NetworkReceiveThroughput | database/network/received_bytes_count | network_bytes_ingress |
| network_transmitted | Bytes/s | NetworkTransmitThroughput | database/network/sent_bytes_count | network_bytes_egress |
| connections | Count | DatabaseConnections | database/network/connections | connection_successful |

In addition, the same metric of a solution may contain necessary information for several vendor-neutral metrics, as shown in Table V, where the same metric (i.e. *api/request_count* and *Transactions* for Google Cloud and Microsoft Azure, respectively) has been mapped to different vendor-neutral metrics. This is due to the fact that the metric outputs contain the necessary information to calculate all the vendor-neutral metrics, that is, the number of requests made for each request type.

For the collection of these vendor-neutral metrics, the multi-cloud broker exports a common API, which must implement all the cloud adapters of the different cloud providers. More specifically, each of these adapters will implement the necessary mapping functions to obtain the vendor-neutral metrics defined in the repository, thus reporting the values for the same vendor-neutral metrics of their services.

As a result of all this diversity and the many specificities, it is not easy to establish a mapping between metrics. To solve this complex problem, we require the knowledge of experts in each of the solutions provided by the different cloud providers in order to define the most complex mappings (indirect mappings) and discover synergies between equivalent solutions. However, the direct mappings that have been established here are broad enough to demonstrate the viability and advantages of this solution, as endorsed by the cloud experts consulted in the course of this research. In fact, the PaaS vendor-neutral metrics repository for each service contains at least one metric for all measurable magnitudes common to all providers.

TABLE VII: MySQL Vendor-Neutral Metrics

| VNM | Unit | AWS Related metric(s) | GCP Related metric(s) | M. Azure Related metric(s) |
|---|---|---|---|---|
| CPU_utilization | % | CPUUtilization | database/cpu/utilization | cpu_percent |
| memory_usage | Bytes | FreeableMemory (Bytes) TotalMemory (GiB) | database/memory/usage | memory_percentmemory_quota |
| memory_utilization | % | FreeableMemory (Bytes) TotalMemory (GiB) | database/memory/utilization | memory_percent |
| disk_usage | Bytes | FreeStorageSpace (Bytes) TotalDisk (GiB) | database/disk/bytes_used | storage_used |
| disk_utilization | % | FreeStorageSpace (Bytes) TotalDisk (GiB) | database/disk/utilization | storage_percent |
| disk_write_ops | Count/s | WriteIOPS | database/disk/write_ops_count | Data Disk Write Operations/Sec OS Disk Write Bytes/Sec |
| disk_read_ops | Count/s | ReadIOPS | database/disk/read_ops_count | Data Disk Read Operations/Sec OS Disk Read Bytes/Sec |
| network_received | Bytes/s | NetworkReceiveThroughput | database/network/received_bytes_count | network_bytes_ingress |
| network_transmitted | Bytes/s | NetworkTransmitThroughput | database/network/sent_bytes_count | network_bytes_egress |
| replica_lag | seconds | ReplicaLag | database/mysql/replication/seconds_behind_\master | seconds_behind_master |
| connections | Count | DatabaseConnections | database/network/connections | connection_successful |

TABLE VIII: Mapping functions that relate the *io_operations* vendor-neutral metric to the corresponding metrics of each of the main cloud providers

| Metric | Unit | AWS | Google Cloud | Microsoft Azure |
|---|---|---|---|---|
| io_operations | operations/s | $= \frac{VolumeReadOps + VolumeWriteOps}{60}$ | $= \frac{operation\_count}{operation\_time}$ | $= Data\ Disk\ ReadOperations/Sec$ $+ DataDiskWriteOperations/Sec$ $+ OSDiskReadOperations/Sec$ $+ OSDiskWriteOperations/Sec$ |

TABLE IX: Mapping functions that relate the *total_requests* vendor-neutral metric to the respective metrics of each of the main cloud providers

| Metric | Unit | AWS | Google Cloud | Microsoft Azure |
|---|---|---|---|---|
| total_requests | count | $= AllRequests$ | $= window(sum, api/request\_count)$ | $= Transactions$ |

### C. Context Metrics

Besides IaaS and PaaS metrics, there is another set of metrics that are common and orthogonal to these services, such as cost or the region where a service is deployed. They have been grouped under the name of context metrics in the broker repository. These metrics are essential for establishing the mapping between SLOs and specific metric conditions, since many of the SLOs that are most important to cloud consumers are related to economic aspects or legislation about the countries where customer data are stored and processed.

Table X shows the current set of context metrics offered by the repository. Accordingly, there are three metrics on the location of services, i.e. *regions*, *countries* and *continents*, which represent the set of regions, countries and continents where there is a deployed service. In addition, there are three other metrics that reflect the cardinality of these sets, representing the total number of different regions, countries and continents where services are currently deployed. Finally, the *total_cost* metric reflects the total cost of all the services, measured in dollars per month.

| Metric | Unit | Description |
|---|---|---|
| regions | - | Set of regions where services are located. $\bigcup_{i=0}^{i=m} regions_i$ |
| countries | - | Set of different countries where services are located. $\bigcup_{i=0}^{i=m} countries_i$ |
| continents | - | Set of different continents where services are located. $\bigcup_{i=0}^{i=m} continents_i$ |
| number_regions | count | card(regions) |
| number_countries | count | card(countries) |
| number_continents | count | card(continents) |
| total_cost | \$/month | Total cost of all the services. $\sum_{i=0}^{i=m} total\_cost_i$ |

TABLE X: Context vendor-neutral metrics

To calculate these metrics, each of the adapters for the different cloud providers export the same metrics, albeit limited to their cloud scope. As a result, the metrics exported

| AWS Region | Context Metric | Value |
|---|---|---|
| us-east-1 | Region | N. Virginia |
| | Country | US |
| | Continent | North America |
| us-east-2 | Region | Ohio |
| | Country | US |
| | Continent | North America |

by the repository are calculated by aggregating the same partial metrics for each of the cloud providers shown in the Description column of Table X. For example, the total cost is the sum of all the total costs exported by the adapters for the different clouds. Similarly, the set of regions is the union of all the sets of regions exported by the adapters for the different clouds. Finally, each adapter has to establish a mapping between the different regions, countries and continents where it is located, as shown in Table XII. In addition, the different regions defined by each cloud provider have been merged into a set of unified regions (vendor-neutral regions). Table XI shows the relationship between the vendor-neutral regions and the regions defined by each cloud provider.

## VII. Evaluation

This section shows how the above solutions to research questions RQ1 and RQ2 work together, and effectively combine and complement each other to provide a solution that demonstrates and verifies hypothesis H1 proposed earlier. To be precise, the evaluation was conducted on two real and typical use cases of multi-cloud environments in different application domains. One use case addressed the IaaS and the other, the PaaS level both within a multi-cloud environment composed of the three leading cloud providers nowadays, i.e. Amazon Web Services or AWS, Google Cloud and Microsoft Azure. This assures a comprehensive validation of the proposed solution. In each use case, the starting point is a SLA representative of the typical needs of cloud consumers in each of the above contexts. Next, we explain the process whereby each SLO that is part of the SLA can be transformed into conditions expressed as vendor-neutral metrics that can be measured and monitored across all the cloud providers.

### A. Multi-cloud publish/subscribe middleware

This first use case proposes the deployment of a publish/subscribe middleware to support an IoT environment that demands high-performance, like smart cities. This is a frequent use case nowadays, since the features offered by the cloud can be exploited by these middlewares to achieve the high performance required by this type of new technological paradigms. For this reason, we selected E-SilboPS, since it is a context-aware content-based publish-subscribe (CBPS) middleware, specially designed to be elastic in cloud environments [13], [28]. To this end, it is composed of four layers of operators (Connection Point, Access Point, Matcher and Exit Point) that can scale independently using the cloud resources provided by the IaaS.

The SLA shown below reflects the requirements of the services that are described in this use case in such contexts.

We find that the SLA is composed of three SLOs: the first establishes the minimum performance of the middleware, measured as notifications/s throughput, the second is related to the maximum total cost of the service per month, and the third imposes a restriction on the geo-location of IaaS resources to ensure that data processing is as close as possible to the place where data are going to be created and consumed (i.e. smart city in Spain).

$$SLO_1 : throughput \geq 30 \; k \; notifications/s$$
$$SLO_2 : total\_cost < 200 \; \$/month$$
$$SLO_3 : region = EU\_Central$$

First, the SLOs that constitute the specified SLA must be translated into objectives expressed through the vendor-neutral set of metrics from the repository (**RQ??**). As explained above, the proposed solution for this purpose is an intelligent knowledge-based system, part of whose knowledge base is shown in Figure 7. We can see the knowledge base is composed of production rules that formally model expert knowledge. Rules R1 to R6 reflect the expert knowledge about the resource-level performance of the four publish/subscribe middlewares, as illustrated in Figure 2. The above study empirically demonstrated that content-based systems or CBPS (i.e. E-SilboPS and BeeTree) were CPU bound, while topic-based systems or TBPS were I/O bound (i.e. ActiveMQ and RabbitMQ). Therefore, these production rules map a throughput SLO in terms of the conditions with respect to maximum CPU or memory usage based on the application domain, i.e. CBPS or TBPS, respectively (N1 & N3).

Note the flexibility of this solution, since all that would be required to support a new publish/subscribe middleware would be to introduce an additional condition to rules R1 or R2, as appropriate. Despite the benefits of the broad coverage offered by this flexibility, it is still possible to express the most specific expert knowledge by restricting the antecedent of the production rule, as illustrated by rule R7 (N2). This rule can be used to introduce concrete E-SilboPS knowledge, since the latency between the different instances of the Matcher operator must be minimal in order to reach high performance (in terms of throughput measured as notifications/s) [13].

On the other hand, production rules R8, R9 and R10 also reflect the expert knowledge referring to the maximum latency between the different cloud provider data centers depending on their geo-location[19]. In this way, the lowest latencies are achieved between data centers in the same region, and they increase depending on whether they are in the same country or the same continent. Therefore, IaaS resources supporting E-SilboPS must be in the same region to ensure minimal latency and achieve the throughput SLO. Finally, rules R11, R12 and R13 do not describe any expert knowledge. They are necessary, however, to add geo-location constraints to vendor-neutral metrics. In the case of the cost metric, the SLO is already expressed as a condition in terms of a vendor-neutral metric, and, therefore, no translation is required.

TABLE XI: Relationship between the vendor-neutral regions and the regions defined by each cloud provider

| vendor-neutral region | AWS | GCP | M. Azure |
|---|---|---|---|
| US_EAST | us-east-1<br>us-east-2 | us-east4<br>us-east1<br>us-central1 | East US<br>East US 2<br>Central US<br>West US 2 |
| US_WEST | us-west-1<br>us-west-2 | us-west1<br>us-west2 | West US<br>West US 2<br>West Central US<br>West Central US |
| CANADA | ca-central-1 | northamerica-northeast1 | Canada Central<br>Canada East |
| SOUTH_AMERICA | sa-east-1 | southamerica-east1 | Brazil South |
| EU_CENTRAL | eu-central-1 | europe-west3<br>europe-west6 | Germany Central<br>Germany Northeast |
| EU_WEST | eu-west-1<br>eu-west-2<br>eu-west-3 | europe-west1<br>europe-west2<br>europe-west4 | West Europe<br>France Central<br>France South |
| EU_NORTH | eu-north-1 | europe-north1 | North Europe<br>UK South<br>UK West |
| AP_NORTH | ap-northeast-1<br>ap-northeast-2<br>ap-northeast-3 | asia-northeast1<br>asia-east1<br>asia-east2 | Japan East<br>Japan West<br>Korea Central<br>Korea South<br>East Asia<br>China East<br>China North<br>China East 2<br>China North 2 |
| AP_SOUTH | ap-southeast-1<br>ap-southeast-2<br>ap-south-1 | asia-southeast1<br>australia-southeast1<br>asia-south1 | Southeast Asia<br>Australia Central<br>Australia Central 2<br>Australia East<br>Australia Southeast<br>West India<br>Central India<br>South India |

TABLE XII: Mapping example of two regions defined by AWS and their respective values in vendor-neutral metrics.

Table XIII gives a step-by-step overview of how, starting from the SLA, the intelligent knowledge-based system is able to map the different SLOs into specific vendor-neutral metric objectives. Each step specifies the facts that have been obtained and the set of production rules that can be activated (conflict set) as their antecedents have been fulfilled. In this case, the inference engine evaluates the rules by forward chaining, applying a conflict resolution strategy that prioritizes the execution matching the most recent facts.

At the end of the process, the set of conditions expressed exclusively by vendor-neutral metrics is output in Step 5 when the conflict set is empty, as shown in Figure 8. We find that the IaaS resources must be deployed in a single region, namely EU_Central, the total cost of all the resources deployed in the different clouds must not be greater than $200 per month, and

CPU use must not exceed 80% of computational resources. Finally, through the above mapping functions, each cloud adapter will monitor and report the necessary resources as vendor-neutral metrics to the multi-cloud broker. In this case, the total cost will be calculated by adding the partial cost reported by each provider. It is important to ensure that no cloud adapter deploys resources outside the EU_Central region at any time and that CPU usage, measured via collectd, does not exceed 80% in any instance.

### B. Multi-cloud RDBMS

The objective of this second use case is to offer a classical OLTP (on-line transactional processing) relational database service (MySQL) as part of a typical three-tier architecture. The requirements of this specific use case, reflected in the SLA of Figure 11, are as follows:

**R1: IF** $App = E - SilboPS \lor App = BeTree$
   **THEN** $AT = CBPS$

**R2: IF** $App = RabbitMQ \lor App = ActiveMQ$
   **THEN** $AT = TBPS$

**R3: IF** $AT = CBPS \land throughput \geq 10 \ k \ notifications/s$
   **THEN** $CPU\_usage \leq 80 \ \%$

**R4: IF** $AT = CBPS \land throughput < 10 \ k \ notifications/s$
   **THEN** $CPU\_usage \leq 90 \ \%$

**R5: IF** $AT = TBPS \land throughput \geq 1 \ k \ notifications/s$
   **THEN** $memory\_usage \leq 40 \ \%$

**R6: IF** $AT = TBPS \land throughput < 1 \ k \ notifications/s$
   **THEN** $memory\_usage \leq 70 \ \%$

**R7: IF** $App = E-SilboPS \land$
   $\land throughput \geq 10 \ k \ notifications/s$
   **THEN** $latency < 30 \ ms$

**R8: IF** $Latency < 30 \ ms$
   **THEN** $number\_regions = 1$

**R9: IF** $30 \ ms \leq latency < 50 \ ms$
   **THEN** $number\_countries = 1$

**R10: IF** $50 \ ms \leq latency < 100 \ ms$
   **THEN** $number\_countinents = 1$

**R11: IF** $continent = X$
   **THEN** $\{X\} \cup continents$

**R12: IF** $country = X$
   **THEN** $\{X\} \cup countries$

**R13: IF** $region = X$
   **THEN** $\{X\} \cup regions$

Fig. 7: Production rules that are part of the knowledge base for the multi-cloud publish/subcribe middleware use case.

$$C_1 : total\_cost < 200 \ \$/month$$
$$C_2 : CPU\_usage \leq 80 \ \%$$
$$C_3 : number\_regions = 1$$
$$C_4 : regions = \{EU\_Central\}$$

Fig. 8: Set of conditions expressed as vendor-neutral metrics for the multi-cloud publish/subscribe middleware use case. The fulfillment of these conditions assures compliance with the SLA.

- The offered service must comply with the European GDPR (General Data Protection Regulation) approved in 2016 by the European Commission [22]. Therefore, data must be stored and processed within Europe ($SLO_1$).
- The average response time of requests to the database

TABLE XIII: Inference process for the translation of SLOs into conditions expressed as vendor-neutral metrics for the multi-cloud publish/subscribe middleware use case

| | Facts | Conflict Set |
|---|---|---|
| **Step 0** | $App = E\text{-}SilboPS$<br>$throughput \geq 30 \ k \ notifications/s$<br>$total\_cost < 200 \ \$/month$<br>$region = EU\_Central$ | $\{\mathbf{R1}, R7, R13\}$ |
| **Step 1** | $App = E\text{-}SilboPS$<br>$throughput \geq 30 \ k \ notifications/s$<br>$total\_cost < 200 \ \$/month$<br>$region = EU\_Central$<br>$AT = CBPS$ | $\{\mathbf{R3}, R7, R13\}$ |
| **Step 2** | $App = E\text{-}SilboPS$<br>$throughput \geq 30 \ k \ notifications/s$<br>$total\_cost < 200 \ \$/month$<br>$region = EU\_Central$<br>$AT = CBPS$<br>$CPU\_usage \leq 80 \ \%$ | $\{\mathbf{R7}, R13\}$ |
| **Step 3** | $App = E\text{-}SilboPS$<br>$throughput \geq 30 \ k \ notifications/s$<br>$total\_cost < 200 \ \$/month$<br>$region = EU\_Central$<br>$AT = CBPS$<br>$CPU\_usage \leq 80 \ \%$<br>$latency < 30 \ ms$ | $\{\mathbf{R8}, R13\}$ |
| **Step 4** | $App = E\text{-}SilboPS$<br>$throughput \geq 30 \ k \ notifications/s$<br>$total\_cost < 200 \ \$/month$<br>$region = EU\_Central$<br>$AT = CBPS$<br>$CPU\_usage \leq 80 \ \%$<br>$latency < 30 \ ms$<br>$number\_regions = 1$ | $\{\mathbf{R13}\}$ |
| **Step 5** | $App = E\text{-}SilboPS$<br>$throughput \geq 30 \ k \ notifications/s$<br>$total\_cost < 200 \ \$/month$<br>$region = EU\_Central$<br>$AT = CBPS$<br>$CPU\_usage \leq 80 \ \%$<br>$latency < 30 \ ms$<br>$number\_regions = 1$<br>$regions = \{EU\_Central\}$ | $\{\}$ |

must not exceed 80 milliseconds to ensure a good quality of service. Furthermore, as this database is part of a three-tier architecture, the maximum number of connections is limited to a maximum of 60 simultaneous connections ($SLO_2$).
- As it is an OLTP service, data availability is essential to ensure business continuity, and therefore availability must be high ($SLO_3$).
- The total cost of the solution must not exceed $300 per month ($SLO_4$).

Some of the production rules that populate the knowledge base of this use case are shown in Figure 10. Rules R1 to R5 model how the GDPR regulation restriction affects service

$SLO_1 : law = GDPR$

$SLO_2 : response\_time < 80\ ms \land max.\ connections = 60$

$SLO_3 : availability = high$

$SLO_4 : total\_cost \leq 300\ \$/month$

Fig. 9: SLOs that constitute the SLA of the OLTP MySQL use case.

deployment, restricting it to European data centers, albeit in three different regions. In the case of an OLAP (on-line analytical processing) application, the number of locations is reduced since high availability is not so critical. Additionally, if there were no regulatory restrictions on data sovereignty, the high availability criterion would be more restrictive since replicas would have to be deployed on more than one continent (N1, N2 & N3).

Rules R6 and R7 describe that the lag between replicas must be limited in order to achieve a specified response time, and these replicas must be deployed in the same region (R8) if a very low lag between replicas is required.

For their part, production rules R9 to R11 formally model the expert knowledge on the computational resource-level behavior of a MySQL OLTP application discovered experimentally (N1 & N3). OLTP tasks are characterized by short-duration transactions with low I/O and CPU usage. Thus, in order to ensure a response time of 80 ms for this type of OLTP database application, the maximum CPU and disk usage thresholds decrease as the maximum number of connections increases, as a consequence of a possibly much more sudden increase in workload due to the greater number of potential requests at any given time. Similarly, due to the more intensive use of CPU and I/O operations that is characteristic of OLAP applications, the production rules would be similar, albeit establishing more restrictive thresholds for CPU usage and the number of I/O operations per second, since these metrics are the KPIs of this type of applications (N2).

As in the previous use case, Table XIV shows the functioning of the intelligent knowledge-based system inference engine, showing the facts and the conflicting set of production rules at each step.

At the end of this the process in Step 4, i.e. when the conflict set is empty and no further production rules can be activated, the set of conditions expressed as vendor-neutral metrics is output (Figure 11). Their fulfillment also ensures compliance with the SLOs. Accordingly, both the regulatory restriction set out in the GDPR and the high availability requirement mean that all MySQL PaaS resources having to be deployed on the European continent, albeit in three different regions. On the other hand, the lag between the different replicas should not be longer than two seconds, and CPU and disk utilization should not exceed 80 %. Note that the vendor-neutral metrics used to express these conditions are either context or PaaS RDBMS (specifically, MySQL) service metrics.

To ensure compliance with this set of conditions, it is essential to be able to measure and monitor the vendor-neutral

TABLE XIV: Inference process for the translation of SLOs in conditions expressed as vendor-neutral metrics for the MySQL OLTP use case

| | Facts | Conflict set |
|---|---|---|
| **Step 0** | $App = MySQL\_OLTP$ <br> $response\_time < 80\ mss$ <br> $max.\ connections = 60$ <br> $total\_cost < 300\ \$/month$ <br> $law = GDPR$ | $\{\mathbf{R1}, R2, R7, R10\}$ |
| **Step 1** | $App = MySQL\_OLTP$ <br> $response\_time < 80\ mss$ <br> $max.\ connections = 60$ <br> $total\_cost < 300\ \$/month$ <br> $law = GDPR$ <br> $number\_continents = 1$ <br> $Europe \in continents$ <br> $regions = \{EU\_NORTH,$ <br> $EU\_WEST, EU\_CENTRAL\}$ | $\{\mathbf{R2}, R7, R10\}$ |
| **Step 2** | $App = MySQL\_OLTP$ <br> $response\_time < 80\ mss$ <br> $max.\ connections = 60$ <br> $total\_cost < 300\ \$/month$ <br> $law = GDPR$ <br> $number\_continents = 1$ <br> $Europe \in continents$ <br> $regions = \{EU\_NORTH,$ <br> $EU\_WEST, EU\_CENTRAL\}$ <br> $number\_regions = 3$ | $\{\mathbf{R7}, R10\}$ |
| **Step 3** | $App = MySQL\_OLTP$ <br> $response\_time < 80\ mss$ <br> $max.\ connections = 60$ <br> $total\_cost < 300\ \$/month$ <br> $law = GDPR$ <br> $number\_continents = 1$ <br> $Europe \in continents$ <br> $regions = \{EU\_NORTH,$ <br> $EU\_WEST, EU\_CENTRAL\}$ <br> $number\_regions = 3$ <br> $replica\_lag \leq 2\ s$ | $\{\mathbf{R10}\}$ |
| **Step 4** | $App = MySQL\_OLTP$ <br> $response\_time < 80\ mss$ <br> $max.\ connections = 60$ <br> $total\_cost < 300\ \$/month$ <br> $law = GDPR$ <br> $number\_continents = 1$ <br> $Europe \in continents$ <br> $regions = \{EU\_NORTH,$ <br> $EU\_WEST, EU\_CENTRAL\}$ <br> $number\_regions = 3$ <br> $replica\_lag \leq 2\ s$ <br> $CPU\_utilization \leq 80\ \%$ <br> $disk\_utilization \leq 80\ \%$ | $\{\}$ |

**R1: IF** $law = GDPR$ **THEN** $Europe \in continents \wedge number\_continents = 1 \wedge$
$\wedge \; regions = \{EU\_NORTH, EU\_WEST, EU\_CENTRAL\}$

**R2: IF** $availability = high \wedge App = MySQL\_OLTP \wedge law = GDPR$ **THEN** $number\_regions = 3$

**R3: IF** $availability = high \wedge App = MySQL\_OLAP \wedge law = GDPR$ **THEN** $number\_regions = 2$

**R4: IF** $availability = high \wedge App = MySQL\_OLTP \wedge \neg \; law = GDPR$ **THEN** $number\_continents = 3$

**R5: IF** $availability = high \wedge App = MySQL\_OLAP \wedge \neg \; law = GDPR$ **THEN** $number\_continents = 2$

**R6: IF** $response\_time \leq 50 \; ms$ **THEN** $replica\_lag \leq 1 \; s$

**R7: IF** $response\_time \leq 100 \; ms$ **THEN** $replica\_lag \leq 2 \; s$

**R8: IF** $replica\_lag \leq 1 \; s$ **THEN** $number\_regions = 1$

**R9: IF** $max. \; connections \leq 50 \wedge App = MySQL\_OLTP \wedge response\_time \leq 100 \; ms$
**THEN** $CPU\_utilization \leq 90 \; \% \wedge disk\_utilization \leq 90 \; \%$

**R10: IF** $50 < max. \; connections \leq 100 \wedge App = MySQL\_OLTP \wedge response\_time \leq 100 \; ms$
**THEN** $CPU\_utilization \leq 80 \; \% \wedge disk\_utilization \leq 80 \; \%$

**R11: IF** $max. \; connections \geq 100 \wedge App = MySQL\_OLTP \wedge response\_time \leq 100 \; ms$
**THEN** $CPU\_utilization \leq 80 \; \% \wedge disk\_utilization \leq 80 \; \%$

Fig. 10: Production rules that form the knowledge base for the MySQL OLTP use case.

$C_1 : number\_continents = 1$

$C_2 : Europe \in continents$

$C_3 : number\_regions = 3$

$C_4 : replica\_lag \leq 2 \; s$

$C_5 : CPU\_utilization \leq 80 \; \%$

$C_6 : disk\_utilization \leq 80 \; \%$

$C_7 : total\_cost < 300 \; \$/month$

Fig. 11: Set of conditions expressed as vendor-neutral metrics. The fulfillment of these conditions assures compliance with the SLA established for the MySQL OLTP use case.

metrics subject to the specified conditions in each of the clouds. To this end, as explained in the proposed solution to RQ2, we have to define the mapping functions that relate these abstract, provider-independent metrics to the metrics exported by each cloud provider. Table XV shows the mapping functions of the vendor-neutral metrics involved in the RDBMS service. We find that the unit and the function whereby each vendor-neutral metric can be calculated is specified based on the metrics exported by each cloud provider (see Table VII). Each of the three cloud providers have equivalent metrics for the *CPU_utilization* and *replica_lag* vendor-neutral metrics, that is, they measure the same magnitude using the same unit and in the same sampling period. The same applies to the *disk_utilization* metric for the Google Cloud and Microsoft Azure providers, where the *database/disk/utilization* and *storage_percent* metrics, respectively, monitor the same magnitude defined in the *disk_utilization* vendor-neutral metric. However, in the case of Amazon Web Services, the mapping function is somewhat more complex, as it must be calculated using the *FreeStorageSpace* and *TotalDisk* metrics. As they are measured in different units, they need to be converted to the same unit before calculating the percentage of disk space used as the complementary value of the percentage of free disk space. On the other hand, the mapping of the context vendor-neutral metrics involved (i.e. *total_cost*, *continents*, *number_continents* and *number_countries*) can be mapped directly as the cumulative value of each of the above metrics collected by each cloud adapter in each cloud, as shown in Section VI-C.

Through these use cases, we were able to evaluate the viability and operation of the solutions proposed to research questions RQ1 and RQ2, checking that they met the stated requirements. Therefore, we were able to validate the original hypothesis H1. For the sake of clarity and understanding, we have only shown the part of the knowledge bases that relevant to each use case. This can be expanded and complicated as required by the expert knowledge. Note also that the vendor-neutral metric conditions output as a result are then evaluated by a system that decides and takes appropriate actions to ensure that they are met. For example, the system may decide to scale out the service in order to distribute the load and decrease the overall CPU usage to ensure compliance with the CPU use conditions, when a resource is close to the defined threshold value. Currently there are a wide variety of such systems implementing purely reactive and even proactive approaches [10], [11]. However, as they are completely decoupled from the solution proposed in this paper, different approaches can be tested in different contexts.

TABLE XV: Mapping functions that relate MySQL OLTP use case vendor-neutral metrics to the corresponding metrics of each of the main cloud providers

| Metric | Unit | AWS | Google Cloud | Microsoft Azure |
|--------|------|-----|--------------|-----------------|
| CPU_utilization | % | $= CPUUtilization$ | $= database/cpu/utilization$ | $= cpu\_percent$ |
| disk_utilization | % | $= \left(1 - \frac{FreeStorageSpace}{2^{30} \times TotalDisk}\right) \times 100$ | $= database/disk/utilization$ | $= storage\_percent$ |
| replica_lag | seconds | $= ReplicaLag$ | $= database/mysql/$ $replication/seconds\_behind\_nmaster$ | $= seconds\_behind\_master$ |

## VIII. Conclusion

In this paper we have reported a solution capable of translating SLAs into specific objectives described as conditions applicable to a set of metrics that can be measured on multiple-cloud (multi-cloud) providers for both IaaS and PaaS services. To this end, the initial hypothesis was divided into two research questions whose solutions complement each other. In response to the first research question, we present a solution based on an intelligent knowledge-based system capable of automatically translating SLAs into specific objectives, the enforcement of which ensures compliance with these SLAs. In this way, the proposed solution serves as an intelligent tutoring system explaining the translations performed, which provides cloud consumers with extra knowledge about the relationship between the performance of their systems and their objectives as expressed in the SLAs. In response to the second research question, we presented a solution that allows the specificities of the different cloud providers to be abstracted by means of a large vendor-neutral metrics repository to monitor not only IaaS and PaaS services but also useful context information, such as service costs and geo-location. This metrics repository is monitored in the different providers of a multi-cloud environment by means of mapping functions that statically relate these vendor-neutral metrics to metrics exported by each cloud provider.

The complementarity of the two solutions results in a system that takes the onus of manually translating its SLAs into lower-level objectives (which tends to be an error-prone process) off the cloud consumer. By expressing these objectives through vendor-neutral cross-provider measurable metrics, all the benefits of multi-cloud can be exploited. This was experimentally verified by evaluating the system based on two use cases in a multi-cloud environment composed of the main cloud providers. In addition, we consulted, as part of this research, cloud experts, and their assessments confirm the validity and completeness of the proposed solution. All these evaluations have proved the viability and correct functioning of the proposed solution, automatically translating different types of SLOs (such as regulatory restrictions with respect to data sovereignty, performance or cost requirements) taking into account the application domain for both IaaS and PaaS services, respectively.

## IX. Future work

Although a lot has been published in the literature on the formalization of SLAs, there is currently no de facto standard for the formalization of these agreements. Due to the importance of SLAs as the root of all the solutions proposed in this paper, it would be valuable to implement and adapt some of these formalization models to the requirements of this particular research, thus defining an interface for integration with other cloud solutions. In addition, the formalization, using ontologies and/or feature models, of the interconnection interface of the two solutions presented in this paper could be very enriching and decouple the two solutions.

Another interesting feature would be to provide the system with a conflict detection and resolution system emulating policy-based solutions [29]. This detection and correction must be carried out at two levels. On the one hand, it is necessary to detect possible conflicts at the SLO level, that is, that no two or more SLOs are contradictory (e.g. $SLO1 = Throughput > 30k\ Notifications/s \wedge SLO2 = Throughput < 30k\ Notifications/s$). The detection of this first case is quite simple, and its correction requires the cloud consumer to redefine any conflicting SLOs. On the other hand, possible conflicts between the objectives of the vendor-neutral metrics output as a result of this second phase must also be detected and corrected. However, they may be harder to detect and resolve. The conflict resolution policy may range widely from the choice of the most restrictive objectives to more complex solutions, such as the prioritization of SLOs.

Another functionality that will add great value to the solution would be to implement a feedback loop to the intelligent knowledge-based system. In this way, the system would acquire knowledge not only from experts but also from the results of system use. The system would refine the production rules based on the outcomes of past experiences, thus becoming a cognitive system.

Finally, some of the services offered by cloud providers are quite complex and require not only theoretical knowledge of the solutions, but also a deeper knowledge that only professional experience can provide. For this reason, we believe that the future collaboration of different cloud service experts can contribute not only to transferring part of their expertise to the intelligent knowledge-based system, but also to collaborative work on the discovery of new relationships between metrics (indirect mappings) in order to create a larger repository of vendor-neutral metrics. On the other hand, several works on cloud computing propose the measurement of new metrics (e.g. energy consumption) that could extend the vendor-neutral metrics repository, increasing the benefits and possibilities of the solution proposed here.

## References

[1] C. Zhu, V. C. M. Leung, L. Shu, and E. C. . Ngai, "Green internet of things for smart world," *IEEE Access*, vol. 3, pp. 2151–2162, 2015.

[2] NIST, "NIST Cloud Computing Standards Roadmap," Tech. Rep., 2013. [Online]. Available: https://nvlpubs.nist.gov/nistpubs/SpecialPublications/NIST.SP.500-291r2.pdf

[3] V. C. Emeakaroha, I. Brandic, M. Maurer, and S. Dustdar, "Low level metrics to high level SLAs - LoM2HiS framework: Bridging the gap between monitored metrics and SLA parameters in cloud environments," *Proceedings of the 2010 International Conference on High Performance Computing and Simulation, HPCS 2010*, pp. 48–54, 2010.

[4] A. Paschke and E. Schnappinger-Gerull, "A Categorization Scheme for SLA Metrics," *Service Oriented Electronic Commerce*, pp. 25–40, 2006. [Online]. Available: http://citeseerx.ist.psu.edu/viewdoc/download?doi=10.1.1.93.1800&rep=rep1&type=pdf

[5] P. Patel, A. Ranabahu, and A. Seth, "Service Level Agreement in Cloud Computing," *Corescholar.Libraries.Wright.Edu*, 2010. [Online]. Available: http://corescholar.libraries.wright.edu/knoesis/78

[6] C. Springs, M. Debusmann, and A. Keller, "Sla-Driven Management of Distributed Systems Using the Common Information Model," *IBM TJ Watson Research Center*, 2003.

[7] C. Quinton, N. Haderer, R. Rouvoy, and L. Duchien, "Towards multi-cloud configurations using feature models and ontologies," *Proceedings of the 2013 international workshop on Multi-cloud applications and federated clouds - MultiCloud '13*, p. 21, 2013. [Online]. Available: http://dl.acm.org/citation.cfm?doid=2462326.2462332

[8] D. Petcu, "Multi-Cloud," *Proceedings of the 2013 international workshop on Multi-cloud applications and federated clouds - MultiCloud '13*, p. 1, 2013. [Online]. Available: http://dl.acm.org/citation.cfm?doid=2462326.2462328

[9] V. Rampérez, J. Soriano, D. Lizcano, and J. A. Lara, "Flas: A combination of proactive and reactive auto-scaling architecture for distributed services," *Future Generation Computer Systems*, vol. 118, pp. 56–72, 2021. [Online]. Available: https://www.sciencedirect.com/science/article/pii/S0167739X20330879

[10] T. Lorido-Botrán, "Auto-scaling techniques for elastic applications in cloud environments," *Technical Report EHU-KAT-IK-09-12 Department*, pp. 1–44, 2012. [Online]. Available: http://www.chinacloud.cn/upload/2012-11/12112312255214.pdf

[11] E. Radhika and G. Sudha Sadasivam, "A review on prediction based autoscaling techniques for heterogeneous applications in cloud environment," *Materials Today: Proceedings*, 2021. [Online]. Available: https://www.sciencedirect.com/science/article/pii/S2214785520394657

[12] L. Rodero-Merino, L. M. Vaquero, V. Gil, F. Galán, J. Fontán, R. S. Montero, and I. M. Llorente, "From infrastructure delivery to service management in clouds," *Future Generation Computer Systems*, vol. 26, no. 8, pp. 1226–1240, 2010. [Online]. Available: http://dx.doi.org/10.1016/j.future.2010.02.013

[13] S. Vavassori, J. Soriano, and Rafael Fernández, "Enabling Large-Scale IoT-Based Services through Elastic Publish/Subscribe," *Sensors (Switzerland)*, 2017.

[14] Y. Chen, S. Iyer, X. Liu, D. Milojicic, and A. Sahai, "Translating service level objectives to lower level policies for multi-tier services," *Cluster Computing*, vol. 11, no. 3, pp. 299–311, 2008.

[15] N. Ferry, A. Rossini, F. Chauvel, B. Morin, and A. Solberg, "Towards model-driven provisioning , deployment , monitoring , and adaptation of multi-cloud systems," *2013 IEEE Sixth International Conference on Cloud Computing*, pp. 887–894, 2013. [Online]. Available: http://www.intercloudtestbed.org/uploads/2/1/3/9/21396364/towards_model-driven_provisioning_deployment_monitoring_and_adaptation_of_multi-cloud_systems.pdf

[16] B. Alattar and N. M. Norwawi, "A personalized search engine based on correlation clustering method," *Journal of Theoretical and Applied Information Technology*, vol. 93, no. 2, pp. 345–352, 2016.

[17] D. Romero, L. Duchien, D. Romero, and L. Duchien, "SALOON : a platform for selecting and configuring cloud environments Cl ´ SA-LOON : a Platform for Selecting and Configuring Cloud Environments Cl ´ HAL Id : hal-01103560," 2015.

[18] FoSII, "FOSII - Foundations of Self-Governing ICT Infrastructures," http://www.infosys.tuwien.ac.at/linksites/FOSII/index.html. [Online]. Available: http://www.infosys.tuwien.ac.at/linksites/FOSII/index.html

[19] D. Petcu, C. Cruaciun, M. Neagul, S. Panica, B. Di Martino, S. Venticinque, M. Rak, and R. Aversa, "Architecturing a Sky Computing Platform," in *Towards a Service-Based Internet. ServiceWave 2010 Workshops*, M. Cezon and Y. Wolfsthal, Eds. Berlin, Heidelberg: Springer Berlin Heidelberg, 2011, pp. 1–13.

[20] D. Zeginis, F. D'Andria, S. Bocconi, J. G. Cruz, O. C. Martin, P. Gouvas, G. Ledakis, and K. A. Tarabanis, "A user-centric multi-paas application management solution for hybrid multi-cloud scenarios," *Scalable Computing*, vol. 14, no. 1, pp. 17–32, 2013.

[21] A. Maarouf, A. Marzouk, and A. Haqiq, "A review of SLA specification languages in the cloud computing," *2015 10th International Conference on Intelligent Systems: Theories and Applications, SITA 2015*, pp. 0–5, 2015.

[22] "Whitepaper on German Data Protection," no. January, 2017.

[23] RabbitMQ, "RabbitMQ - Messaging that just works," http://www.rabbitmq.com/. [Online]. Available: https://www.rabbitmq.com/

[24] Apache Software Foundation, "Apache ActiveMQ," http://activemq.apache.org/. [Online]. Available: http://activemq.apache.org/

[25] M. Sadoghi and H.-a. Jacobsen, "BE-Tree: An index structure to efficiently match Boolean expressions over high-dimensional discrete space," *SIGMOD '11: Proceedings of the 2011 ACM SIGMOD International Conference on Management of Data*, pp. 637–648, 2011.

[26] E. A. Feigenbaum, "Expert systems in the 1980s," *State of the art report on machine intelligence. Maidenhead: Pergamon-Infotech*, 1981.

[27] B. K. Bose, "Expert System, Fuzzy Logic, and Neural Network Applications in Power Electronics and Motion Control," *Proceedings of the IEEE*, vol. 82, no. 8, pp. 1303–1323, 1994.

[28] V. Ramperez, J. Soriano, and D. Lizcano, "A Multidomain Standards-Based Fog Computing Architecture for Smart Cities," *Wireless Communications & Mobile Computing*, vol. 2018, 2018.

[29] N. Dunlop, J. Indulska, and K. Raymond, "Methods for Conflict Resolution in Policy-Based Management Systems," *7th Int. Enterprise Distributed Object Comp. Conf. (EDOC)*, pp. 98–111, 2003. [Online]. Available: http://dx.doi.org/10.1109/EDOC.2003.1233841